\documentclass[twoside]{LCWS11}
\usepackage[latin1]{inputenc}
\usepackage[dvips]{graphicx,epsfig,color}
\usepackage{wrapfig,rotating}
\usepackage{amssymb,amsmath,array}
\usepackage{cite}

\def \ptgg  {$p_T^{\gamma\gamma}$}
\def \mgg  {$M_{\gamma\gamma}$}
\def \dphigg {$\Delta\phi_{\gamma\gamma}$}

\begin{document}
\title{
QCD measurements at the Tevatron} 
\author{Dmitry Bandurin \\ ({\it for the D0 and CDF Collaborations})
\vspace{.3cm}\\
Florida State University,  Department of Physics, \\
Tallahassee, FL 32306, USA
\vspace{.1cm}\\
}

\maketitle

\begin{abstract}
Selected quantum chromodynamics (QCD) measurements
performed at the Fermilab Run II Tevatron $p\bar{p}$ collider 
running at $\sqrt{s} = 1.96$ TeV by CDF and D0 Collaborations are presented.
The inclusive jet, dijet production and three-jet cross section measurements
are used to test perturbative QCD calculations,
constrain parton distribution function (PDF) determinations, and extract a precise value of
the strong coupling constant, $\alpha_s(m_Z) = 0.1161^{+0.0041}_{-0.0048}$.
Inclusive photon production cross-section
measurements reveal an inability of next-to-leading-order (NLO) perturbative QCD (pQCD) calculations
to describe low-energy photons arising directly in the hard scatter. 
The diphoton production cross-sections check the validity
of the NLO pQCD predictions, soft-gluon resummation methods 
implemented in theoretical calculations, and contributions from 
the parton-to-photon fragmentation diagrams.
Events with $W/Z$+jets productions are used to measure many kinematic
distributions allowing extensive tests and tunes of predictions from 
pQCD NLO and Monte-Carlo (MC) event generators. 
%
The charged-particle transverse momenta ($p_T$) and multiplicity distributions 
in the inclusive minimum bias events are used to tune non-perturbative 
QCD models, including those describing the multiple parton interactions (MPI).
Events with inclusive production of $\gamma$ and 2 or 3 jets 
are used to study increasingly important MPI phenomenon at high $p_T$,
measure an effective interaction cross section, $\sigma_{\rm eff} = 16.4\pm 2.3$ mb,
and limit existing MPI models.
\end{abstract}

\section{Introduction}

QCD, the theory of the strong interaction between quarks and
gluons, is heavily tested in experimental studies at hadron colliders.
QCD results from the CDF and D0 collaborations obtained with integrated luminosity
up to 8 fb$^{-1}$ are reviewed in this paper.
These results provide a crucial tests for pQCD, PDFs, the strong coupling constant, 
non-perturbative models describing parton fragmentation, and MPI phenomena.
At the same time, the results are used to search for new phenomena
and impose limits on the corresponding models.
The performed extensive studies prepare a solid base for the LHC era of $pp$ collisions.

\section{Jet production}

Thorough testing of pQCD at short distances 
is provided through measurements of differential inclusive jet, dijet and three-jet cross
sections. The measurements of the inclusive jet cross sections done by the D0 \cite{incj_d0} and CDF \cite{incj_cdf}
collaborations are in agreement with pQCD predictions in a few jet rapidity regions.
However, data with uncertainties lower than theoretical (mostly PDF) ones, favor a smaller
gluon content at high Feynman $x$ ($>$0.2).
The D0 inclusive jet data has also been used to extract values of the strong coupling constant
$\alpha_s$ in the interval of $50 < p_T^{\rm jet} < 145$ GeV \cite{d0_as}.
The best fit over 22 data points leads 
to $\alpha_s(m_Z)=0.1161^{+0.0041}_{-0.0048}$ with improved accuracy from the Run I CDF result
\cite{cdf_as} and also in agreement with result from HERA jet data \cite{hera_as}.

The inclusive dijet cross sections have been measured in the D0 \cite{dijets_d0} and CDF \cite{dijets_cdf}
experiments. Both measurements cover the mass range up to about 1.2 TeV with good agreement with pQCD, 
and no indication of any new physics.
CDF imposed limits on some models with exotic particles decaying into two jets \cite{dijets_cdf}.
D0 results are compared to  {\sc mstw2008} \cite{mstw} and {\sc cteq6.6m} \cite{cteq} PDF sets. 
They are in a better agreement
with {\sc mstw2008}  and are systematically lower than the central pQCD prediction at high rapidities
($1.2<|y|<2.4$).

\begin{figure}[htbp]
\hspace*{0mm}  \includegraphics[scale=0.85]{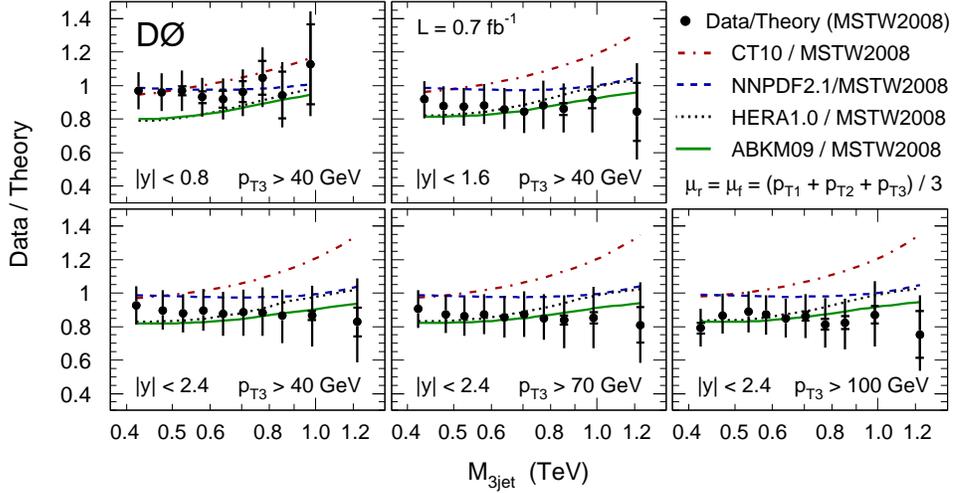}
\caption{The 3-jet mass differential cross section in the three jet rapidity intervals are
compared to NLO pQCD with different PDF sets.}
\label{fig:3jm}
\end{figure}

The D0 collaboration has also measured the three-jet mass cross sections  using jets with leading (in $p_T$) 
jet $p_T>150$ GeV, and considering three regions with different lower cut on the 3rd jet $p_T$ ($40,70,100$ GeV)
and three different jet rapidity regions ($|y|<2.4,1.6,0.8$) \cite{3jets_d0}.
Results are shown in Fig.~\ref{fig:3jm} and in agreement with NLO pQCD predictions,
which use {\sc mstw2008}, {\sc ct10} \cite{CT10}, {\sc nnpdf2.1} \cite{nnpdf}, {\sc hera1.0} \cite{hera10} and 
{\sc abkm09} \cite{abkm09} PDF sets.
Results favor more {\sc mstw2008} and {\sc nnpdf2.1}  PDFs. 
D0 has also presented a measurement of the ratio of the inclusive 3-jet to 2-jet production cross sections \cite{r32}.
The shape of the ratio is well described by NLO QCD and is 
practically independent of the PDF set. The results can potentially be used to test the running of $\alpha_s$ up to
a $p_T$ scale of 500 GeV.

\begin{figure}[htbp]
\hspace*{30mm}  \includegraphics[scale=0.13]{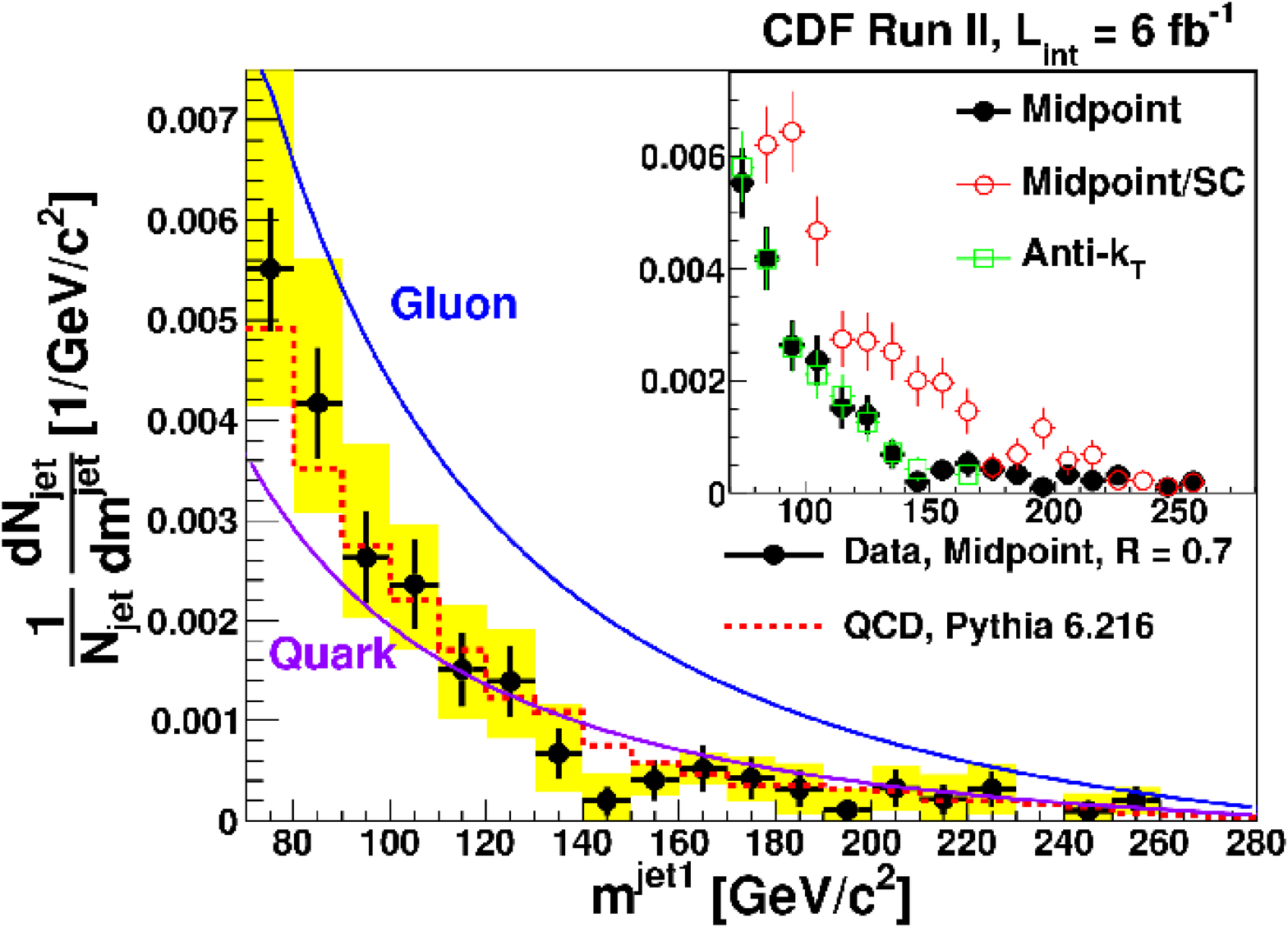}
\caption{
Distribution of jet mass 
for jets with $R=0.7$; black crosses are data, red dashed is QCD MC, 
theoretical ``all quarks'' and ``all gluons'' curves are presented as well; the inset plot compares 
the results with Midpoint/SC and Anti-kT jet algorithms.}
\label{fig:jets_struct}
\end{figure}

The CDF collaboration studied structure of high $p_T$ jets by selecting only events with at least one jet
having $p_T>400$ GeV, $0.1<|y|<0.7$ and considering jets with cone sizes $R=0.4,0.7$ and $1.0$ \cite{cdf_jet_struct}.
Such studies can be used to tune parton showering and search for heavy resonances decaying hadronically.
The jet mass is calculated using 4-vectors of calorimeter towers in a jet. 
Fig.~\ref{fig:jets_struct}
shows the jet mass distribution for $R=0.7$ at high masses.
The data are in agreement {\sc pythia} predictions and interpolate between  
the QCD LLA predictions \cite{jet_struct_th}  for quark and gluon jets,
and confirm that the high mass jets are mostly caused by quark fragmentation.

\begin{figure}[htbp]
\hspace*{0mm}  \includegraphics[scale=0.4]{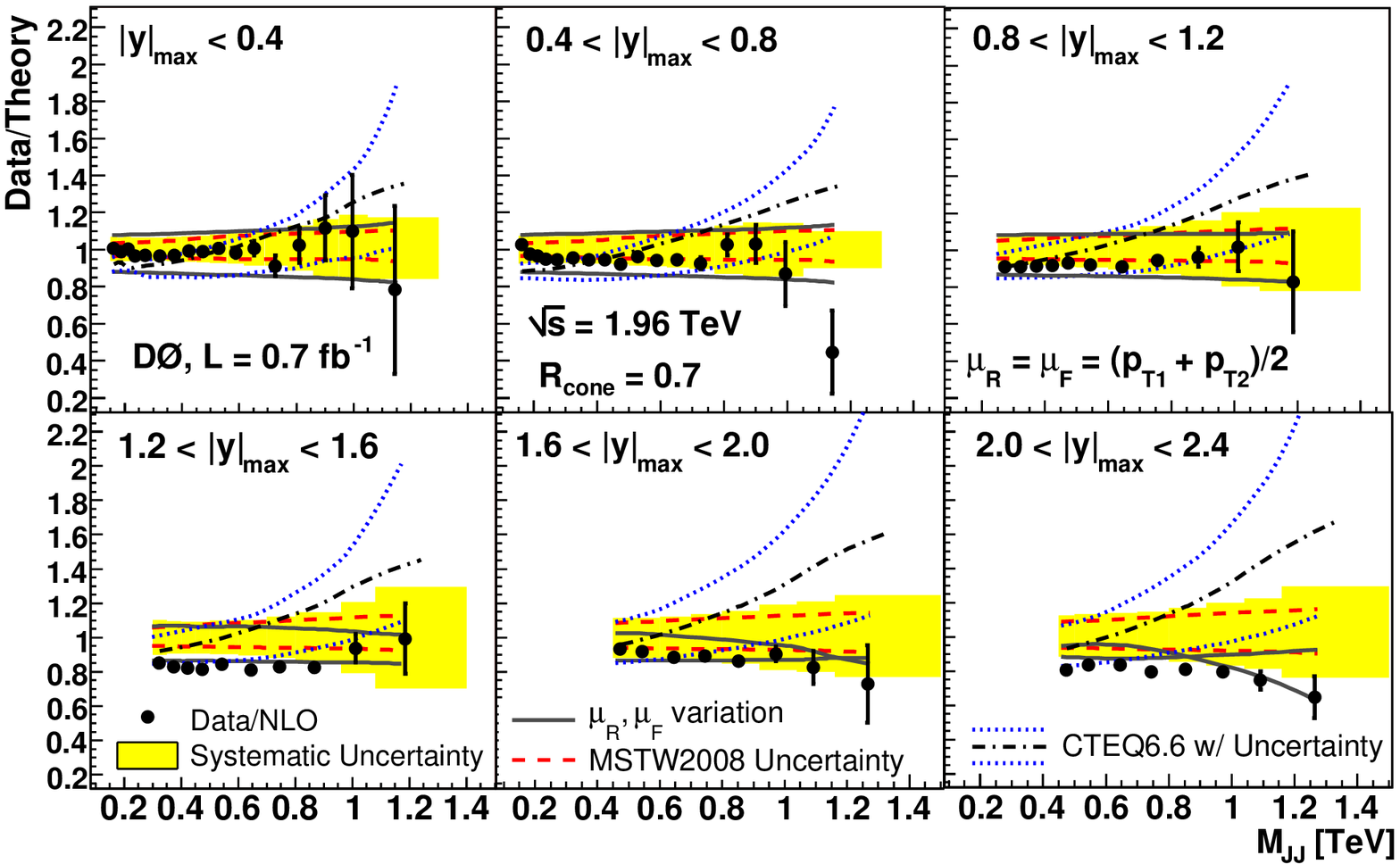}
\hspace*{0mm}  \includegraphics[scale=0.12]{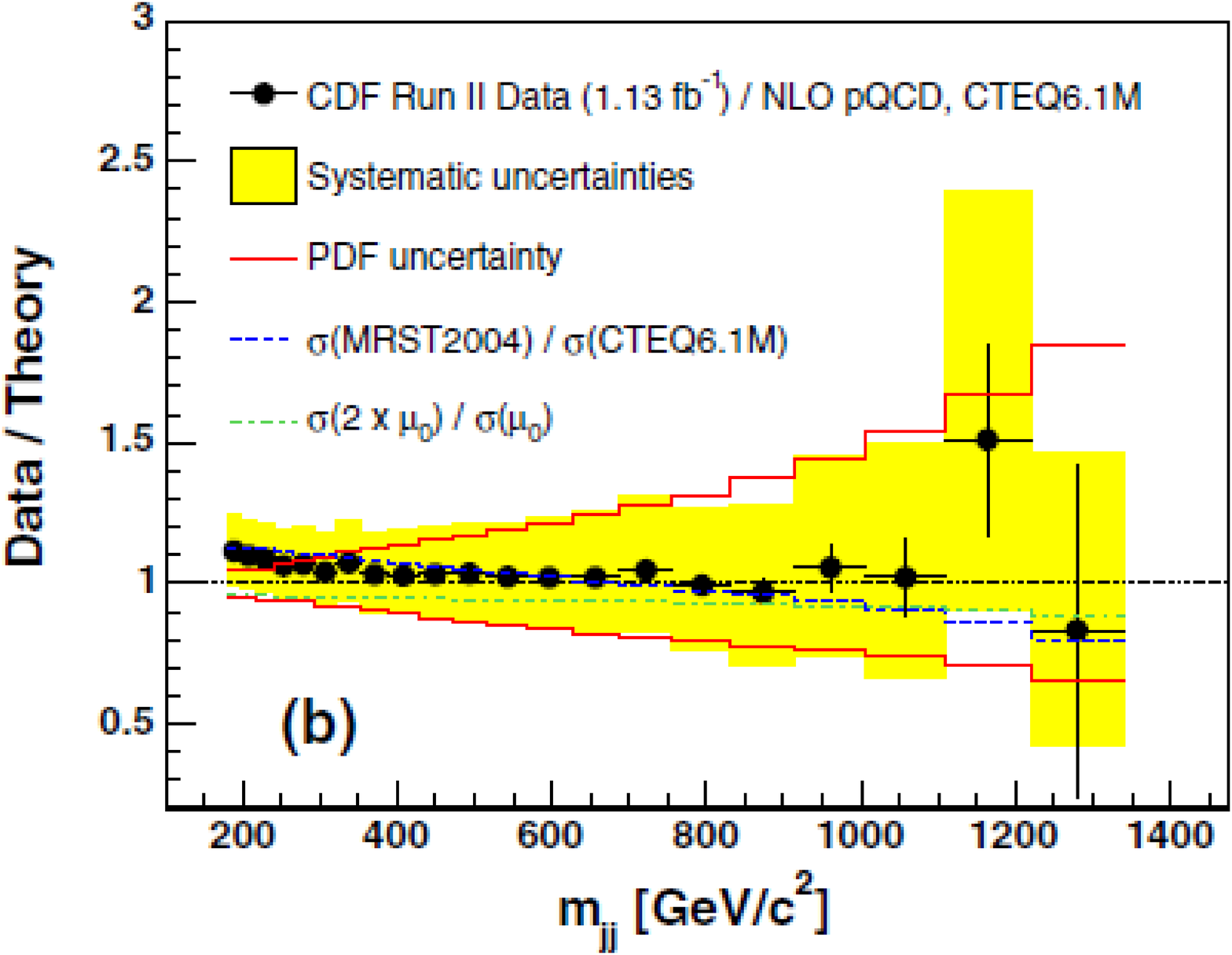}
\caption{Ratios of data/theory for the dijet mass cross sections measured in D0 and CDF are shown on the left
and right plots.}
\label{fig:dij_m}
\end{figure}

\section{$W/Z$ + jets production}

Both collaborations have extensively studied the $W/Z$ + jet productions since these events
are the main background to top-quark, Higgs boson, SUSY and many other new physics production channels.
In this section we review some of the latest results.

\begin{figure}[htbp]
\hspace*{0mm}  \includegraphics[scale=0.15]{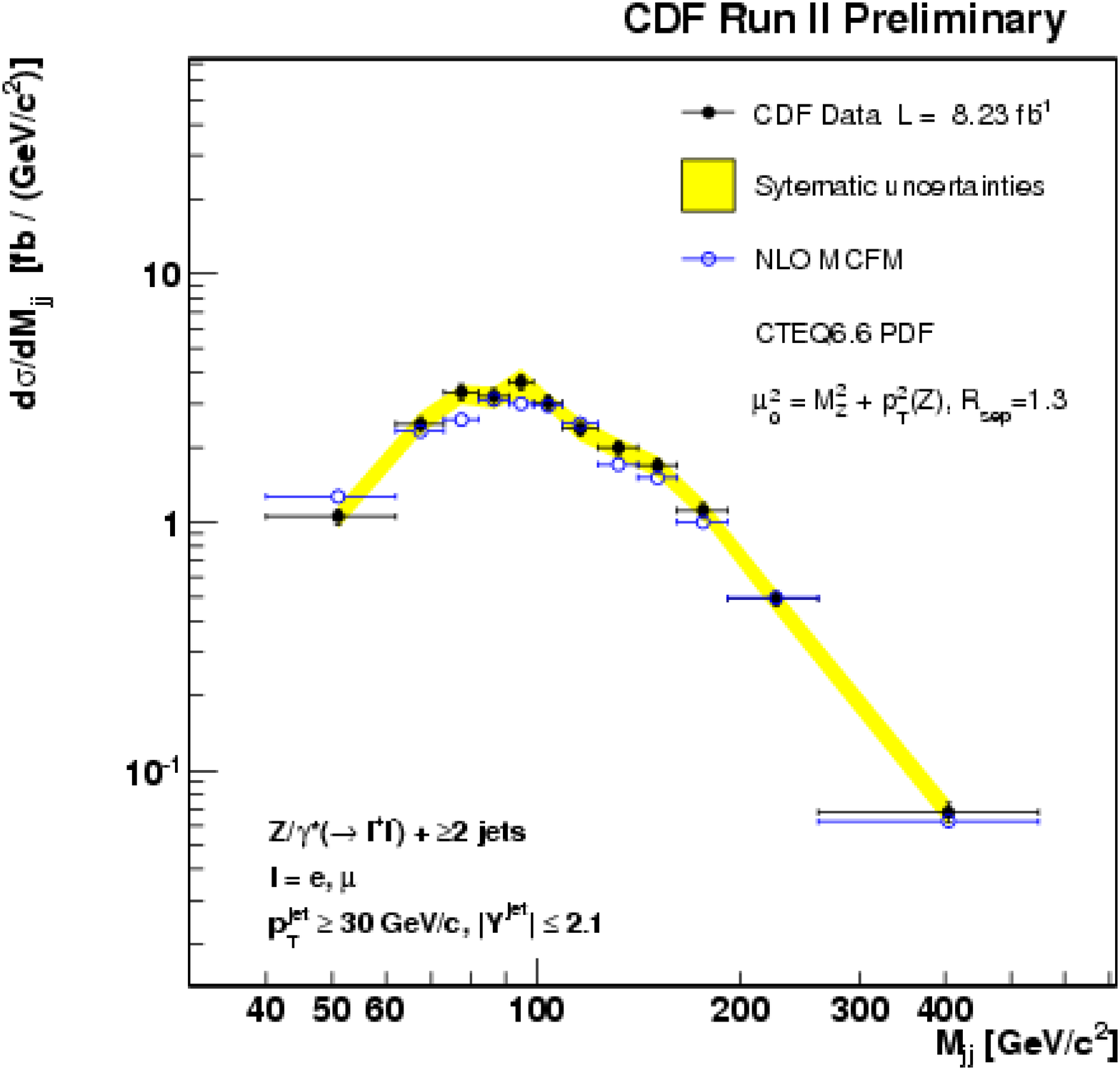}
\hspace*{0mm}  \includegraphics[scale=0.15]{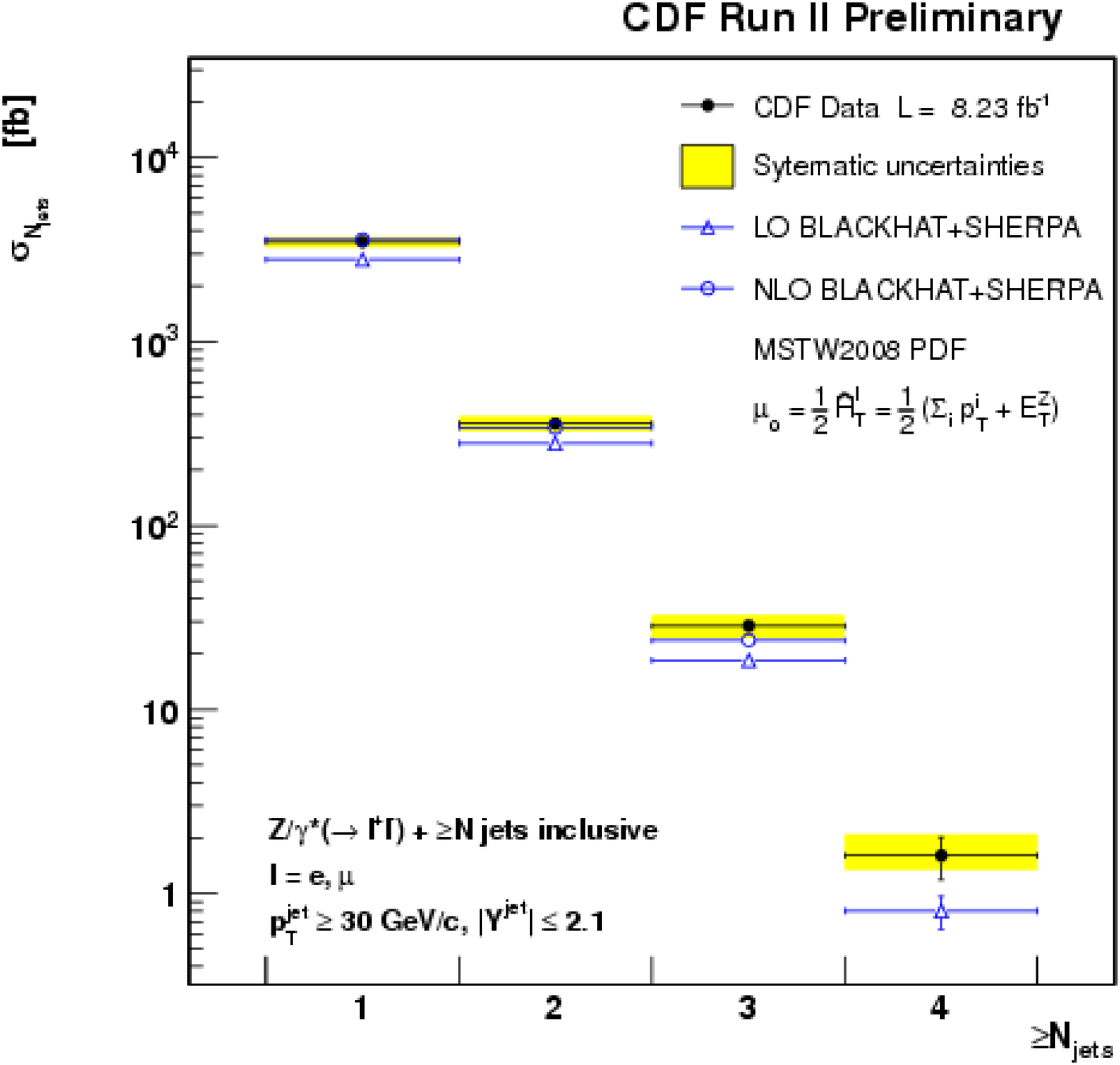}
\caption{Left: measured inclusive cross section for $Z/\gamma^*$+jets production as a function 
of dijet mass compared to NLO pQCD predictions as determined using MCFM. 
Right: measured cross section as a function of inclusive jet multiplicity compared to LO and NLO pQCD predictions as determined using MCFM. 
}
\label{fig:zj_cdf}
\end{figure}

Fig.~\ref{fig:zj_cdf} shows the inclusive cross section for $Z/\gamma^\ast$+jets production measured
by CDF \cite{cdf_zj} as a function of dijet mass and jet multiplicity. The measurements are 
compared to LO and NLO pQCD predictions obtained with MCFM \cite{mcfm} and are in good agreement
with the NLO theory predictions.
D0 measured jet $p_T$ inclusive cross sections of $W+n$-jet production for jet multiplicities $n=1-4$ \cite{d0_wj}.
The measurements are compared to the NLO predictions for $n=1-3$
and to LO predictions for $n = 4$. The measured cross sections are generally found to
agree with the NLO calculation although certain regions
of phase space are identified where the calculations could be improved.

D0 recently published the measured cross section ratio 
$\sigma(Z+b)/\sigma(Z+{\rm jet})=0.0193\pm0.0022({\rm stat})\pm0.0015({\rm syst})$
for events with jet $p_T>20$ GeV and $|\eta|<2.5$ \cite{d0_zb}. 
This most precise measurement of the $Z+b$ fraction is consistent with the NLO theory prediction, $0.0192\pm0.0022$,
(done with MCFM, renormalization and factorization scales set at $m_Z$) and 
the CDF result 
\cite{cdf_zb}.
the CDF collaboration measured the cross section of $W+b$-jet production 
$\sigma(W+b)\cdot Br(W\to l\nu) = 2.74\pm 0.27({\rm stat})\pm0.42({\rm syst})$ pb
with jet $p_T>20$ GeV, $|\eta|<2.0$ and $l=e,\mu$. 
The measurement significantly exceeds the NLO prediction $1.2\pm0.14$ pb.

\begin{figure}[htbp]
\vspace*{-10mm}
\hspace*{0mm}  \includegraphics[scale=0.31]{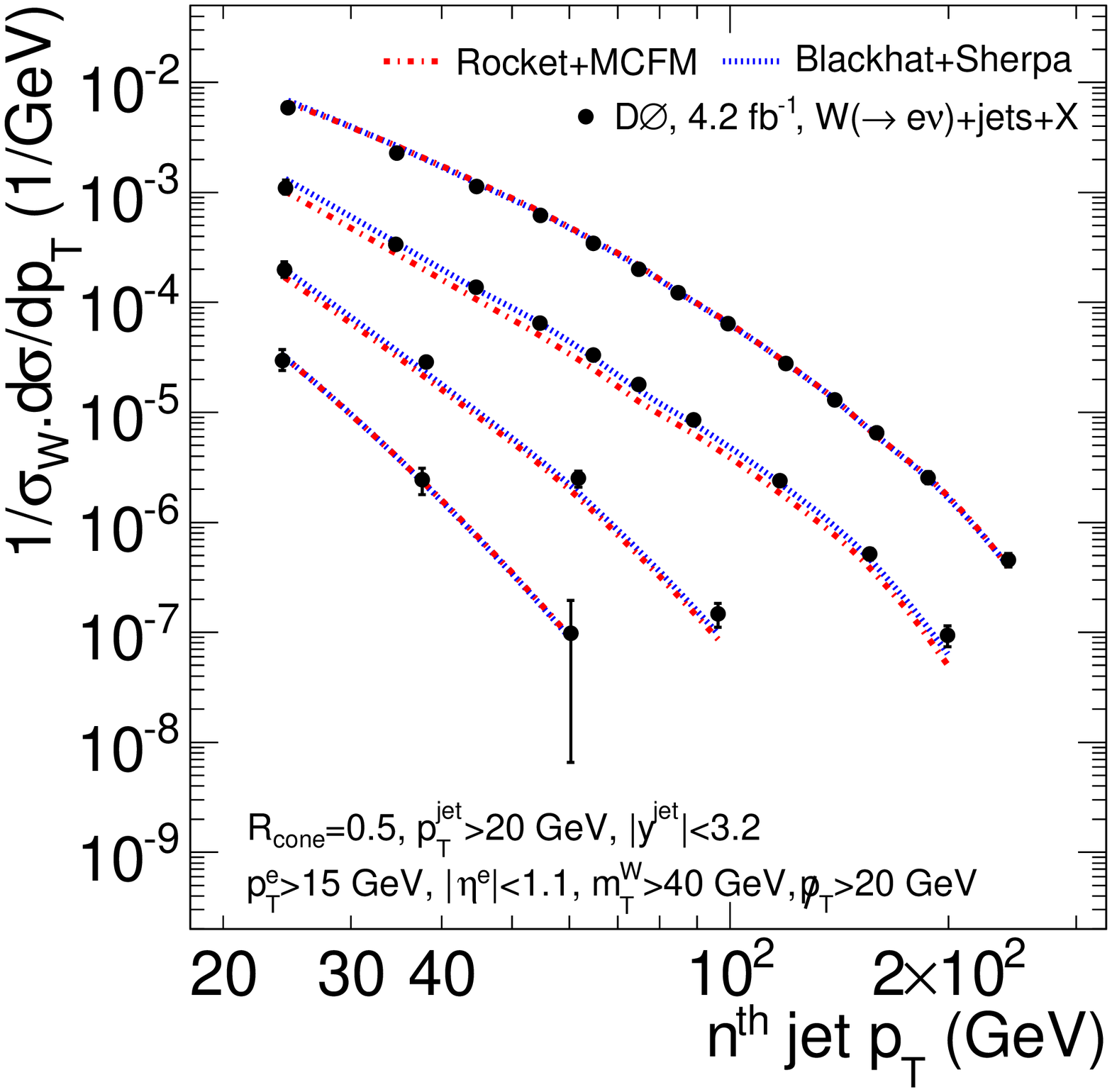}
\hspace*{7mm}  \includegraphics[scale=0.29]{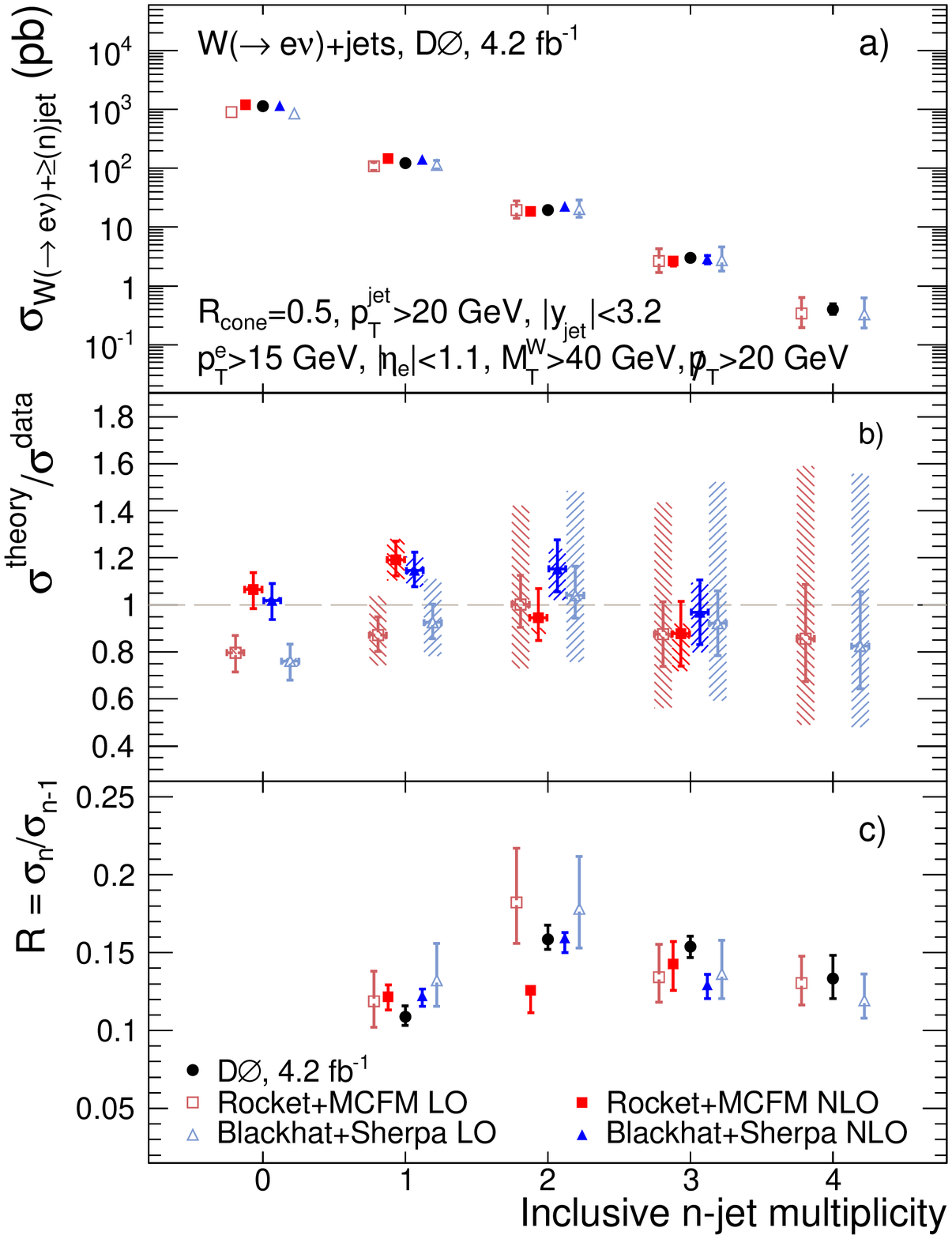}
\caption{Left: measured $W +n$ jet differential cross section as a function of jet 
$p_T$ for $n=1-4$, normalized to the inclusive $W\to e\nu$ cross section. The $W +1$ jet inclusive spectra
are shown by the top curve, the $W +4$ jet inclusive spectra by
the bottom curve. The measurements are compared to the
fixed-order NLO predictions for $n=1-3$ and to LO predictions for $n = 4$.
Right: (a) total inclusive $n$-jet cross sections 
$\sigma_n$ 
as a function of $n$, (b) the ratio of the theory predictions to the measurements, 
and (c) $\sigma_n/\sigma_{n-1}$ ratios for data,
Blackhat+Sherpa and Rocket+MCFM. 
The hashed areas represent the theoretical uncertainty arising from the choice of
renormalization and factorization scale.}
\label{fig:wj_d0}
\end{figure}
\begin{figure}[htbp]
\hspace*{0mm}  \includegraphics[scale=0.28]{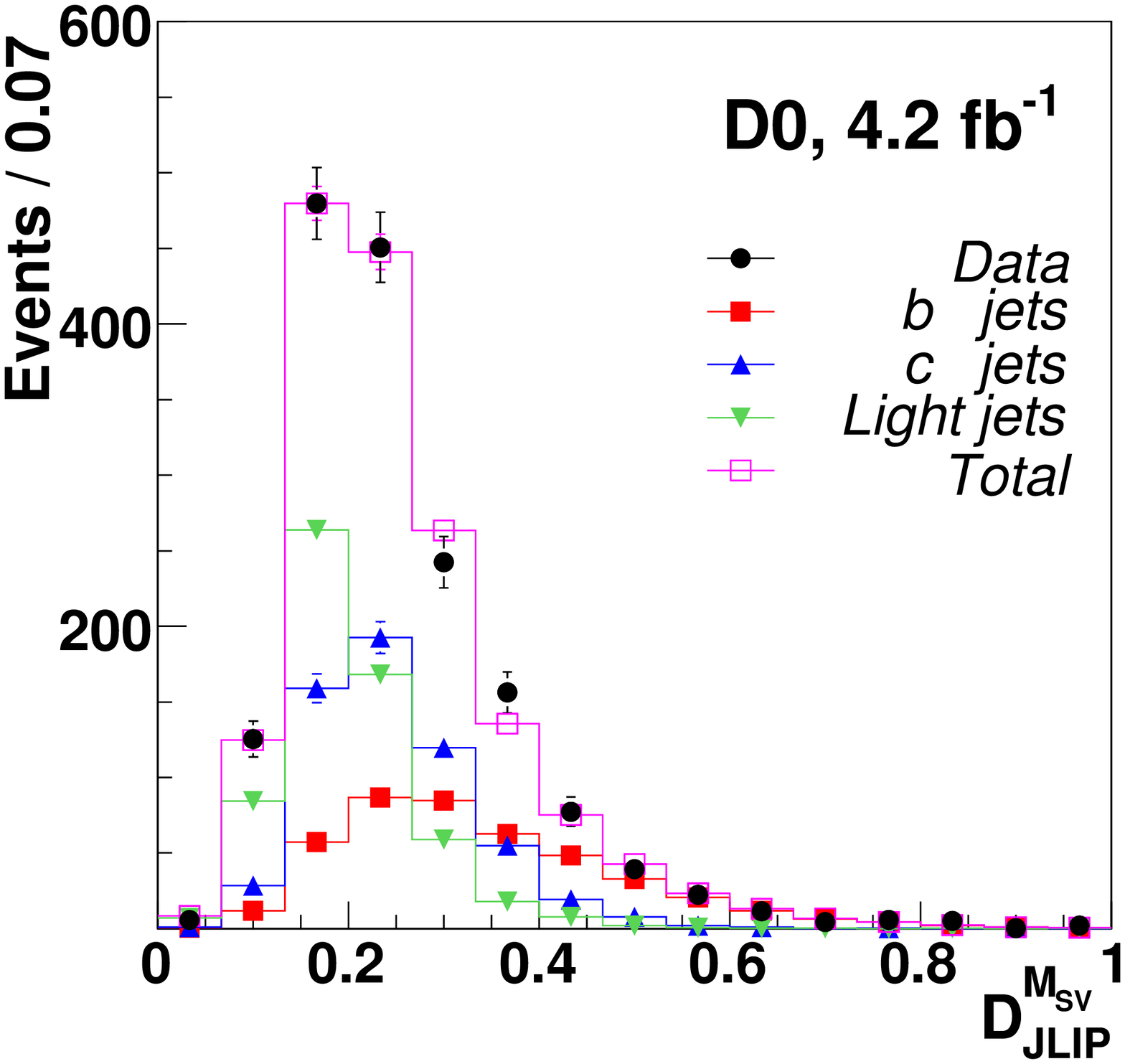}
\hspace*{5mm}  \includegraphics[scale=0.28]{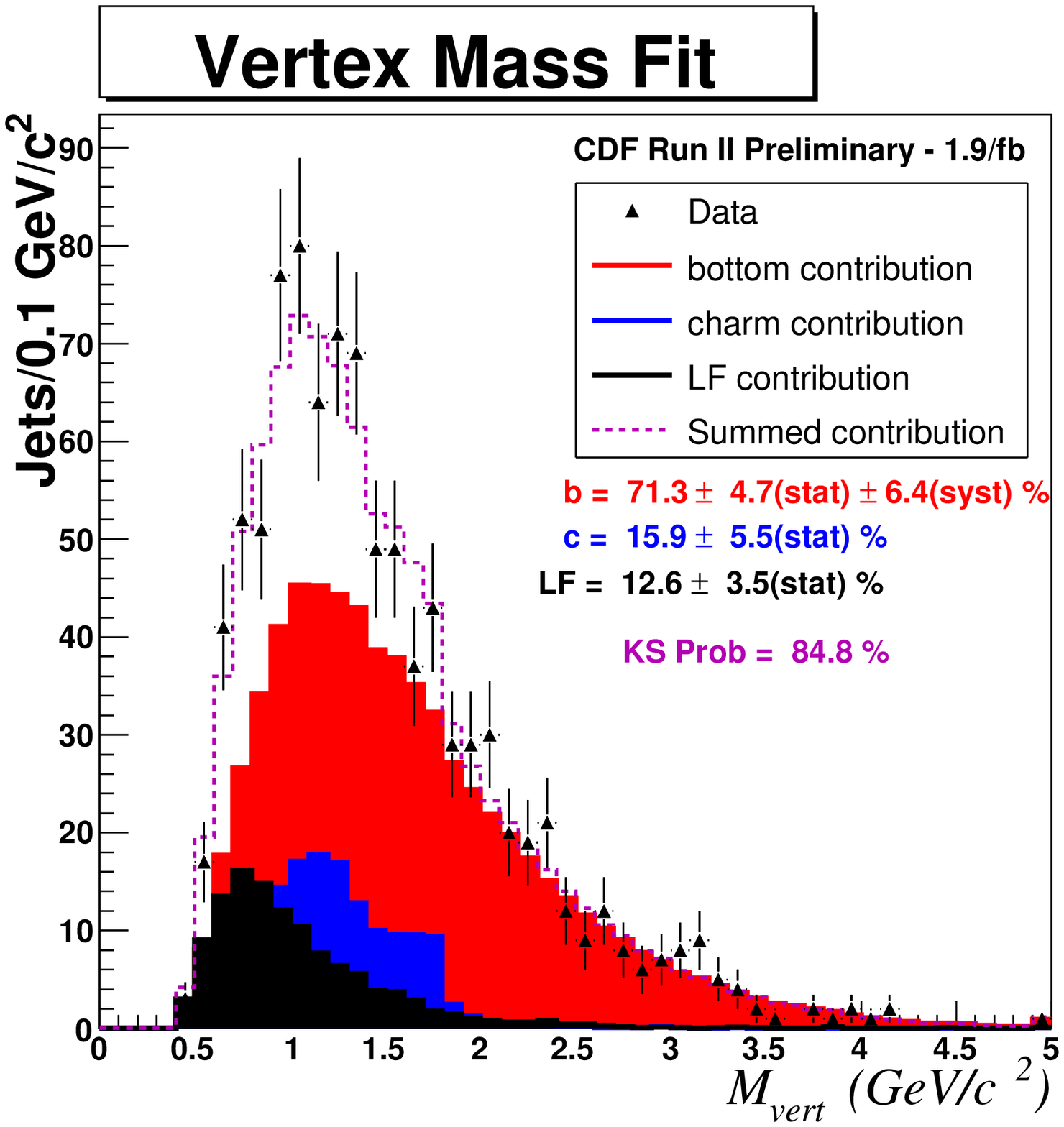}
\caption{Left (D0):  
the distributions of the $b$, $c$, light jets and data over the $b-$jet discriminant;
MC templates are weighted by the fractions found from the fit to data. 
Right (CDF): the secondary vertex mass fit for the tagged jets in the selected sample.} 
\label{fig:wzb}
\end{figure}

\section{Photon production}

Since high $p_T$ photons emerge directly from $p\bar{p}$ collisions and provide
a direct probe of the parton hard scattering dynamics, they are of permanent interest in
high energy physics. The inclusive photon production cross sections measured by D0 and CDF
in the central rapidity region \cite{d0_incgam,cdf_incgam} 
are in agreement within experimental uncertainties and both indicate a difficulty for NLO pQCD
to describe the low $p_T$ behavior.

\begin{figure}[htbp]
\hspace*{-2mm}  \includegraphics[scale=0.23]{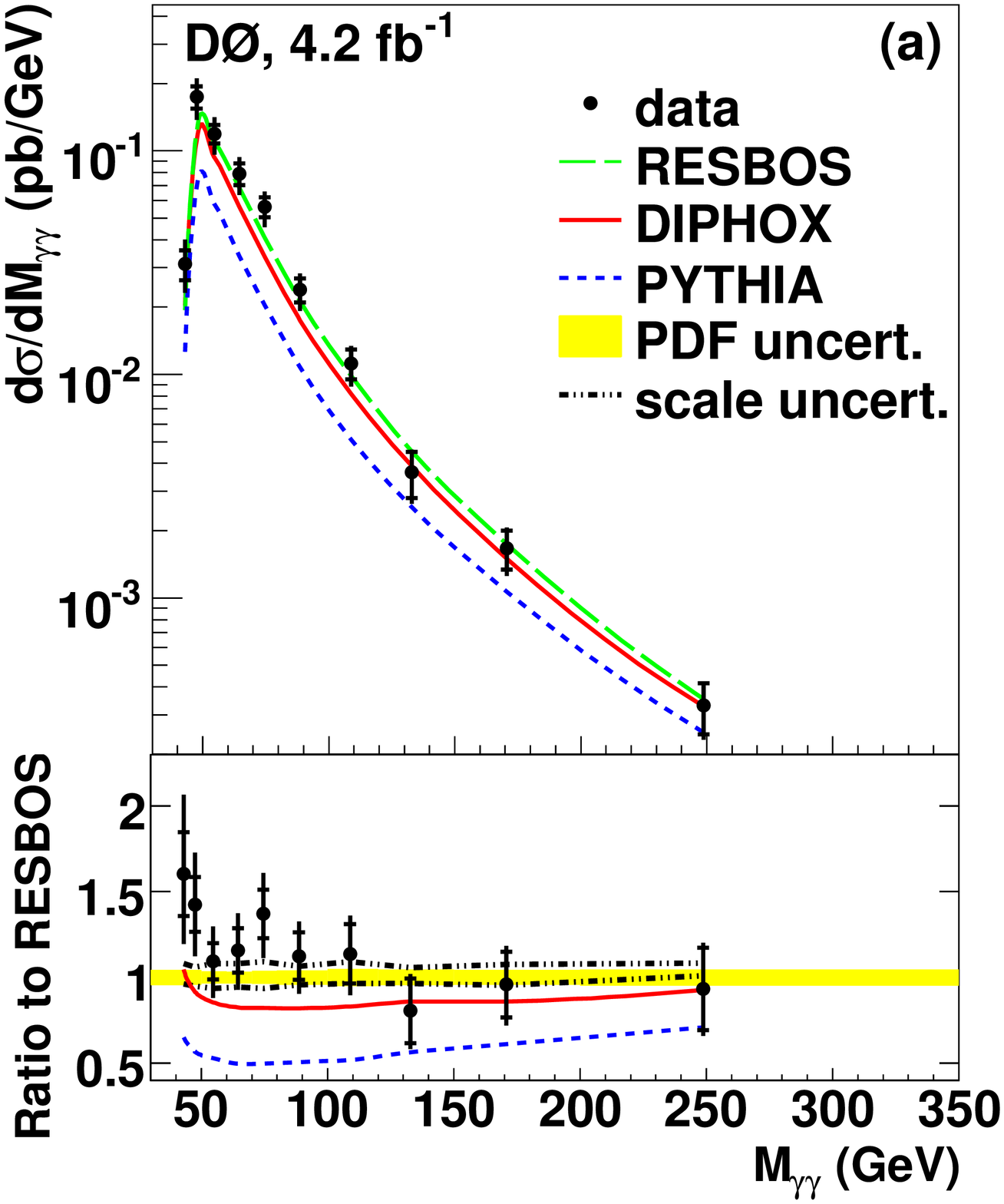}
\hspace*{0mm}  \includegraphics[scale=0.23]{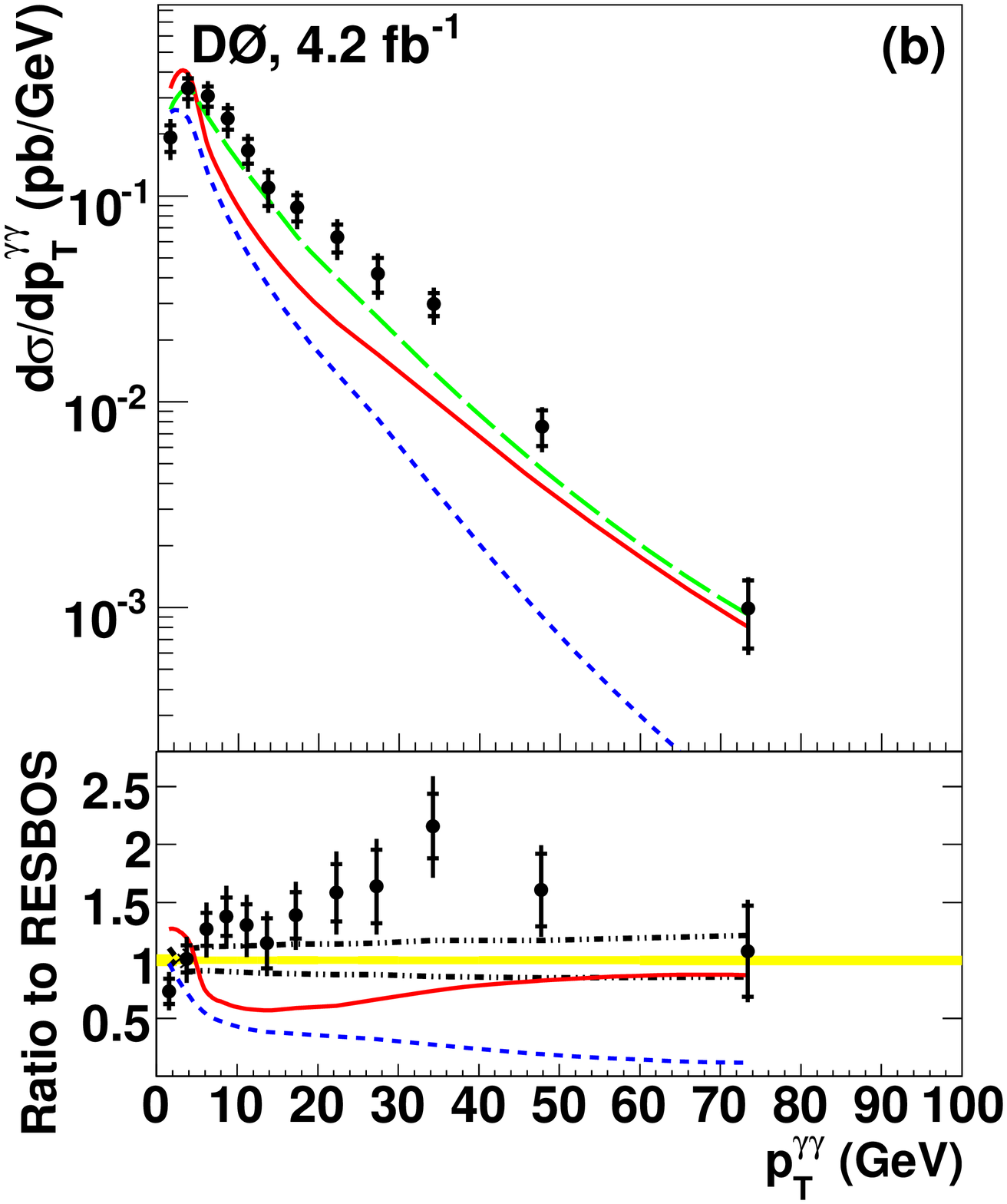}
\hspace*{0mm}  \includegraphics[scale=0.23]{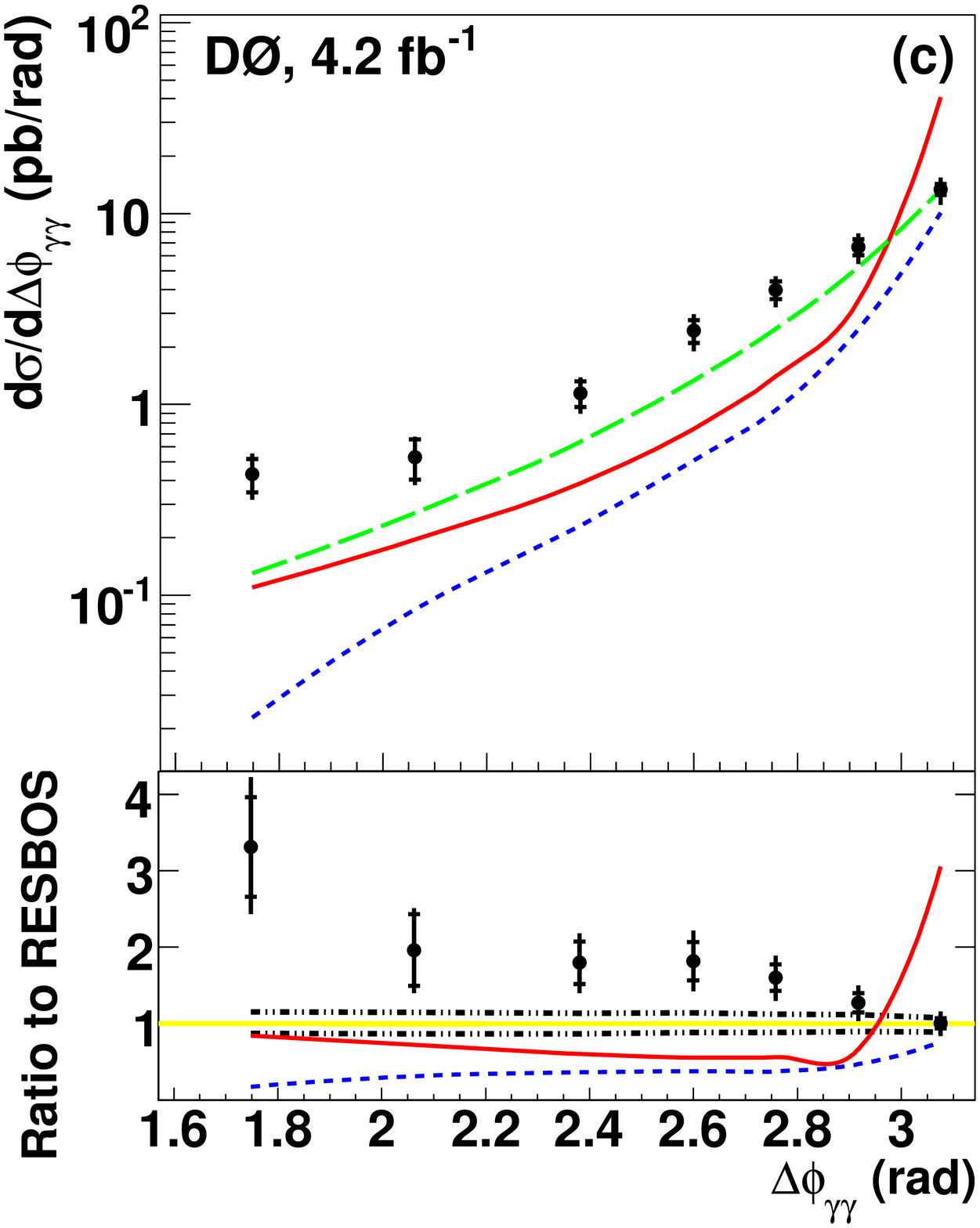}
\caption{The measured differential diphoton production cross sections as functions of
(a) \mgg, (b) \ptgg and (c) \dphigg ~in D0 experiment. 
The data are compared to the theoretical predictions from
{\sc resbos}, {\sc diphox}, and {\sc pythia}.
The ratio  of differential cross sections between data and {\sc resbos} are displayed as black points
with uncertainties in the bottom plots.
The solid (dashed) line shows the ratio of the predictions from {\sc diphox} ({\sc pythia}) to those from {\sc resbos}.
In the bottom plots, the scale uncertainties are shown by dash-dotted lines and the PDF uncertainties by shaded regions.} 
\label{fig:diph_d0}
\end{figure}

\begin{figure}[htbp]
\hspace*{-1mm}  \includegraphics[scale=0.22]{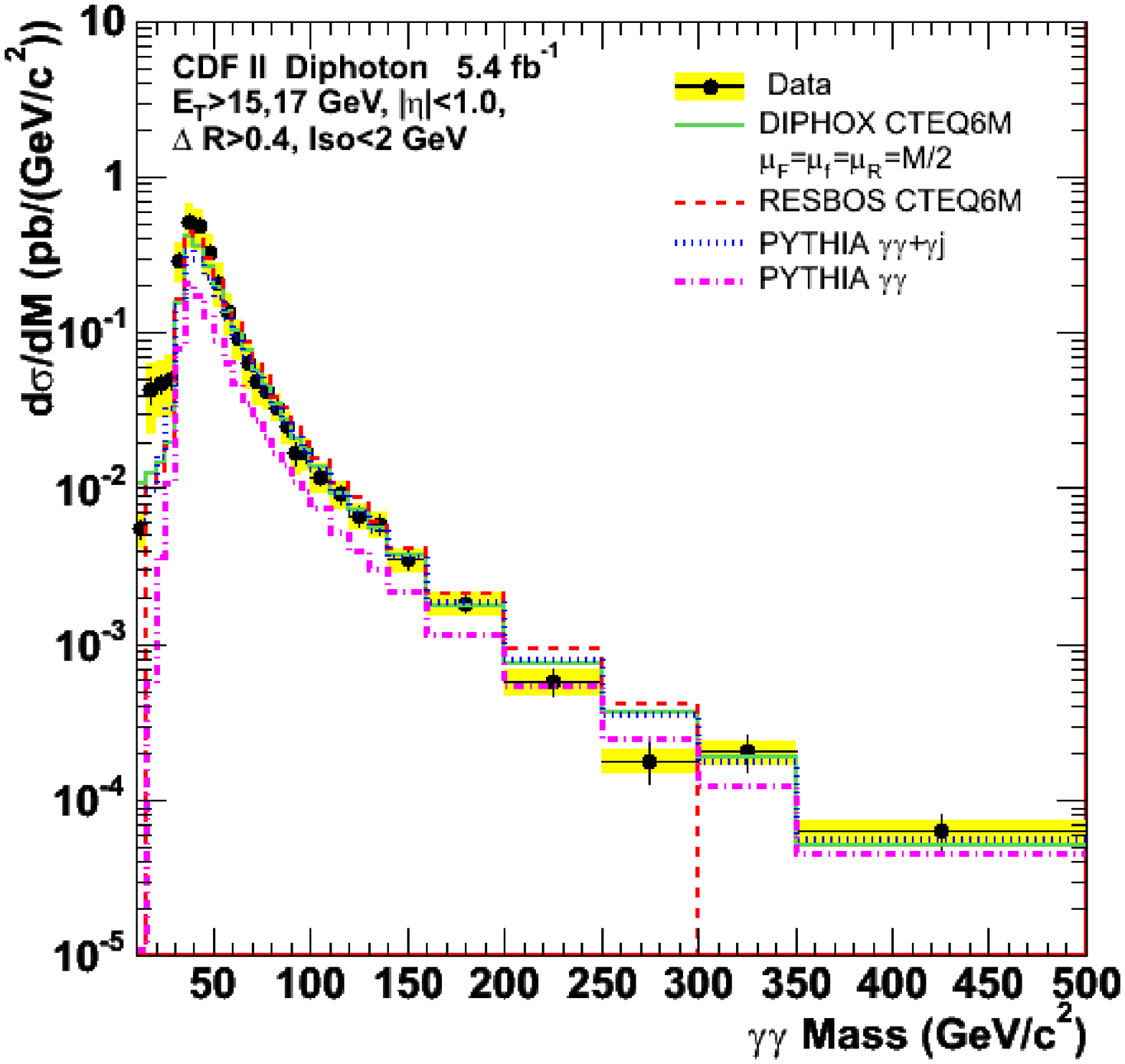}
\hspace*{-2mm}  \includegraphics[scale=0.22]{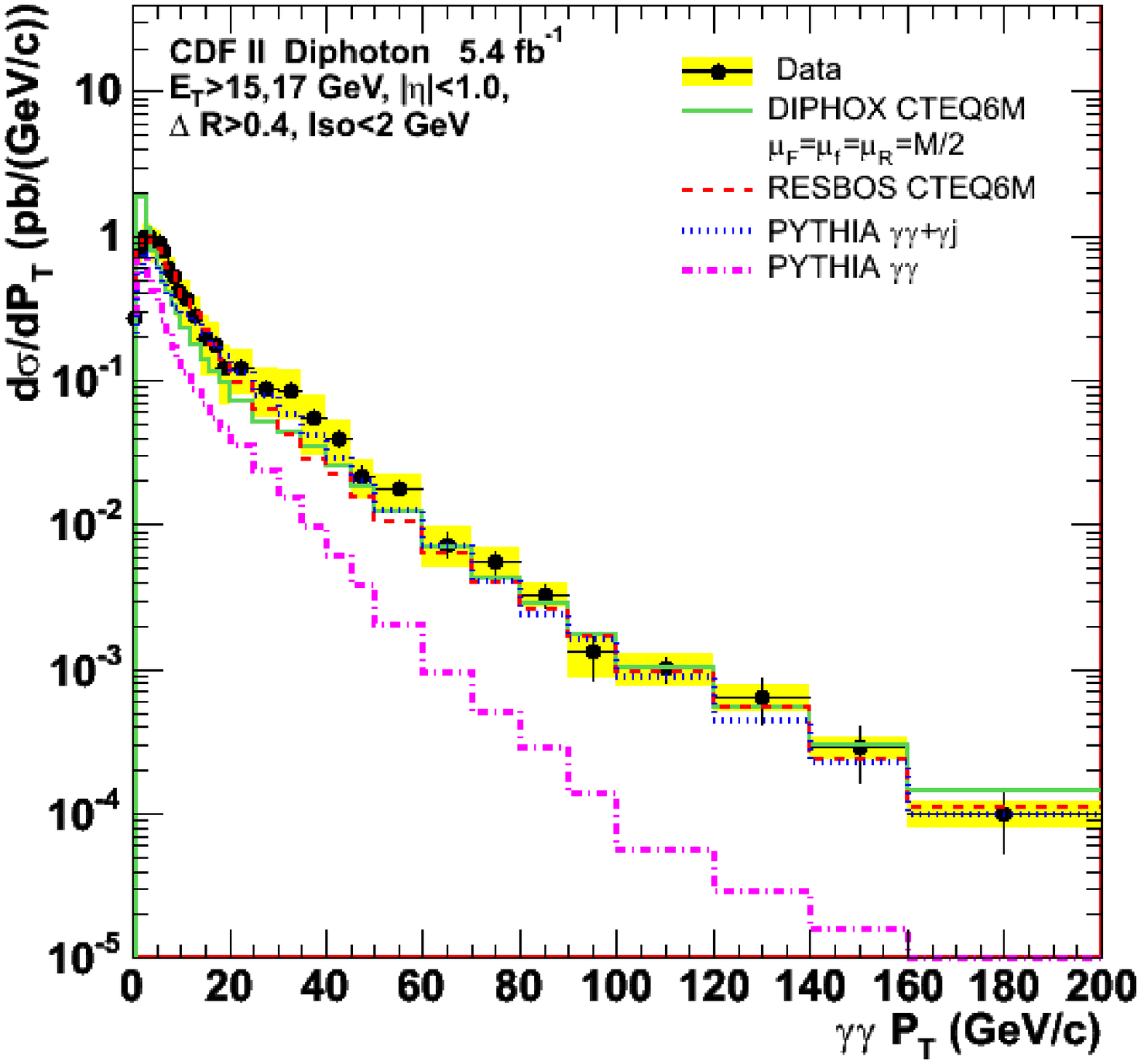}
\hspace*{-2mm}  \includegraphics[scale=0.22]{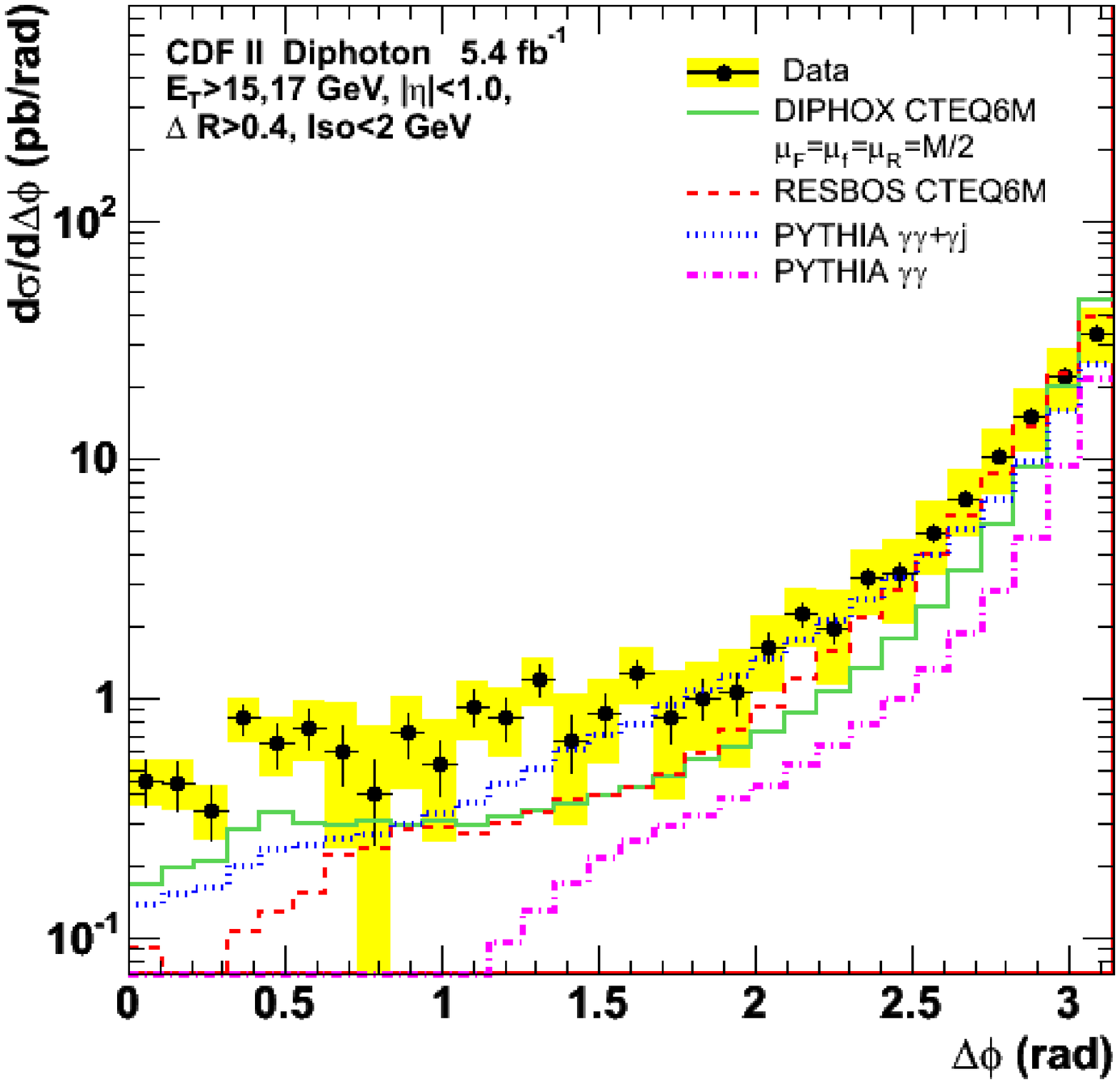}
\caption{The measured differential diphoton production cross sections as functions of
(from left to right) \mgg,  \ptgg and \dphigg ~in CDF experiment.} 
\label{fig:diph_cdf}
\end{figure}

In light of the Higgs boson search and other possible resonances decaying to a photon pair,
both the collaborations performed a thorough study of the diphoton production.
D0 measured the diphoton cross sections (see Fig.~\ref{fig:diph_d0}) as a function of the diphoton
mass \mgg, the transverse momentum of the diphoton system \ptgg, the azimuthal angle
between the photons \dphigg, and the polar scattering angle of the photons.
The latter three cross sections are
also measured in the three \mgg ~bins, $30-50, 50-80$ and $80-350$ GeV.
The measurements are compared to NLO QCD 
and {\sc pythia} \cite{pythia-ref} predictions.
The results show that the largest discrepancies between data and NLO predictions 
for each of the kinematic variables originate from the lowest \mgg~region
(\mgg~$<50$~GeV), 
where the contribution from $gg\to\gamma\gamma$ is expected to be largest \cite{diphoton_d0}. 
The discrepancies between data and the theory predictions are
reduced in the intermediate \mgg~region, and a quite satisfactory description
of all kinematic variables is achieved for the \mgg$>80$~GeV region, the
relevant region for the Higgs boson and new phenomena searches.
The CDF collaboration measured the diphoton production cross sections
functions of \mgg,  \ptgg and \dphigg. They are shown in Fig.~\ref{fig:diph_cdf}.
None of the models describe the data well in all kinematic regions, in particular
at low diphoton mass (\mgg$<60$ GeV), low \dphigg ($<1.7$ rad) and moderate
\ptgg ($20-50$ GeV).

\section{Multiple parton interactions}

\begin{figure}[htbp]
\hspace*{5mm}  \includegraphics[scale=0.32]{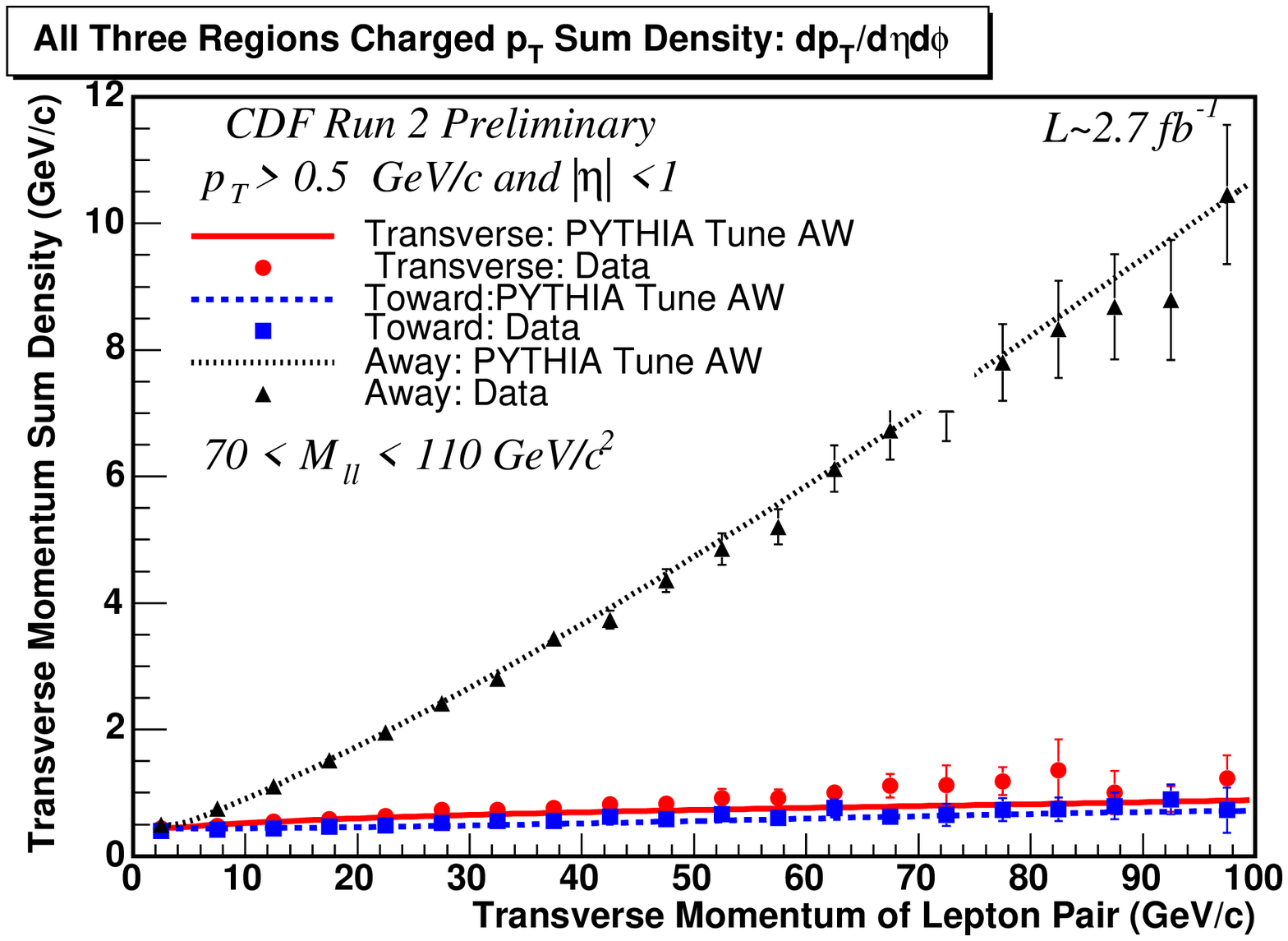}
\hspace*{0mm}  \includegraphics[scale=0.32]{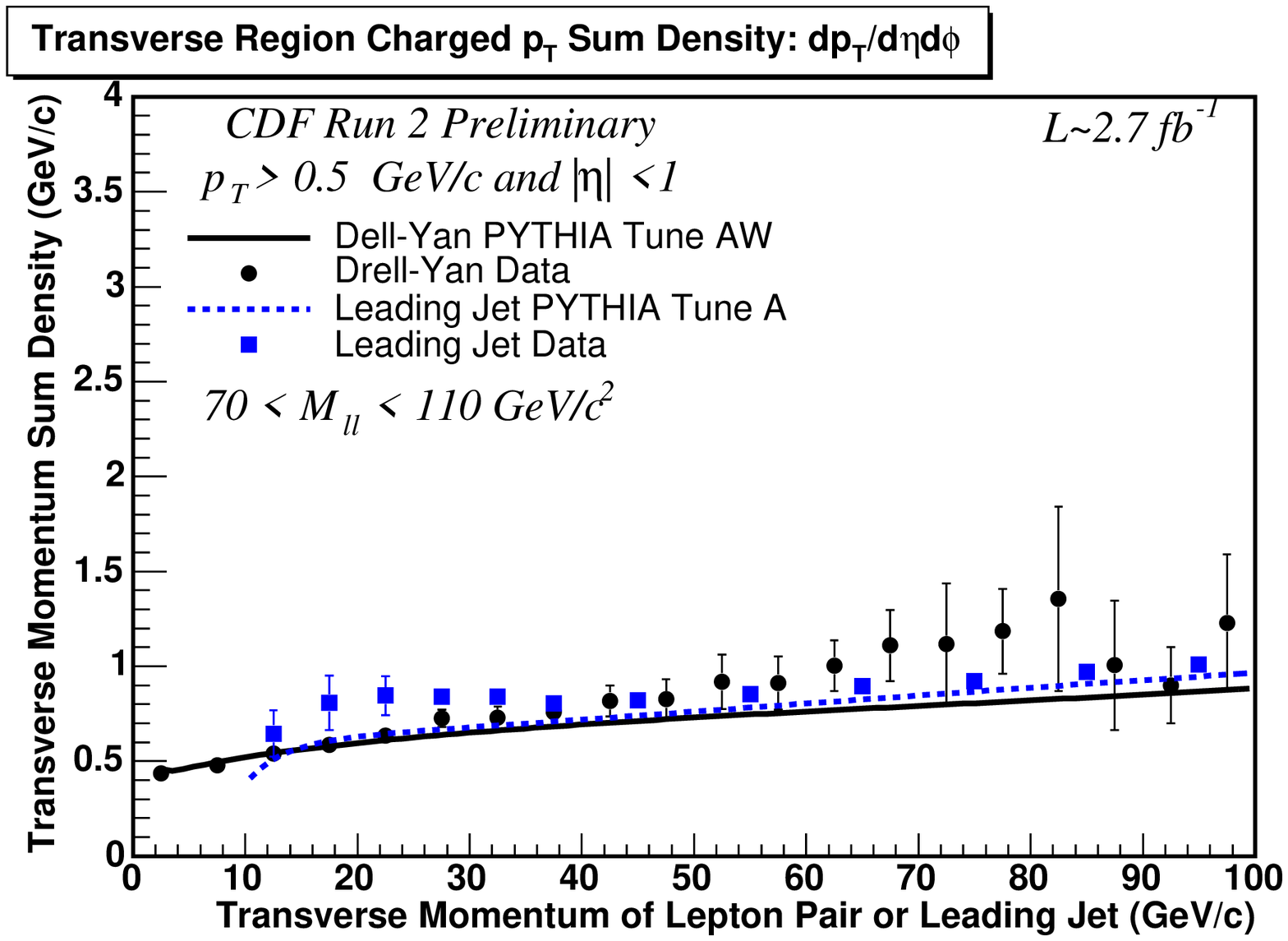}
\caption{Left: the comparison of the total charged-particle $p_T$ sum density $dp_T/d\eta d\phi$ 
in the three regions  in the Drell-Yan events: ``transverse'', ``away'' and ``toward''. 
Right: the $p_T$ sum density in the transverse region in Drell-Yan and jet production
as a function of the lepton pair or leading jet $p_T$.} 
\label{fig:ptdens_cdf}
\end{figure}

The CDF and D0 collaborations comprehensively studied the phenomenon of MPI events in a few Run II measurements.
In this section we mention some of the recently published results.
CDF studied charged-particle $p_T$ sum densities and multiplicities in Drell-Yan 
and jet events \cite{mpi_cdf}. Specifically, both distributions have been analyzed in the three
regions: towards the total lepton pair (Z-boson) $\vec{p}_T$, opposite to this direction (``away'' region)
and in the region transverse to the lepton pair/jet $\vec{p}_T$.
The charged-particle $p_T$ sum density in the three regions in the Drell-Yan events  is shown 
on the left plot of Fig.~\ref{fig:ptdens_cdf} as a function of the lepton pair $p_T$.
The same quantity is also plotted for the transverse region on the right plot. 
One can see a similar trend in both Drell-Yan and jet events which can be considered
as MPI universality. The tuned {\sc pythia} describes the data very well.

\begin{figure}[htbp]
\hspace*{35mm}  \includegraphics[scale=0.32]{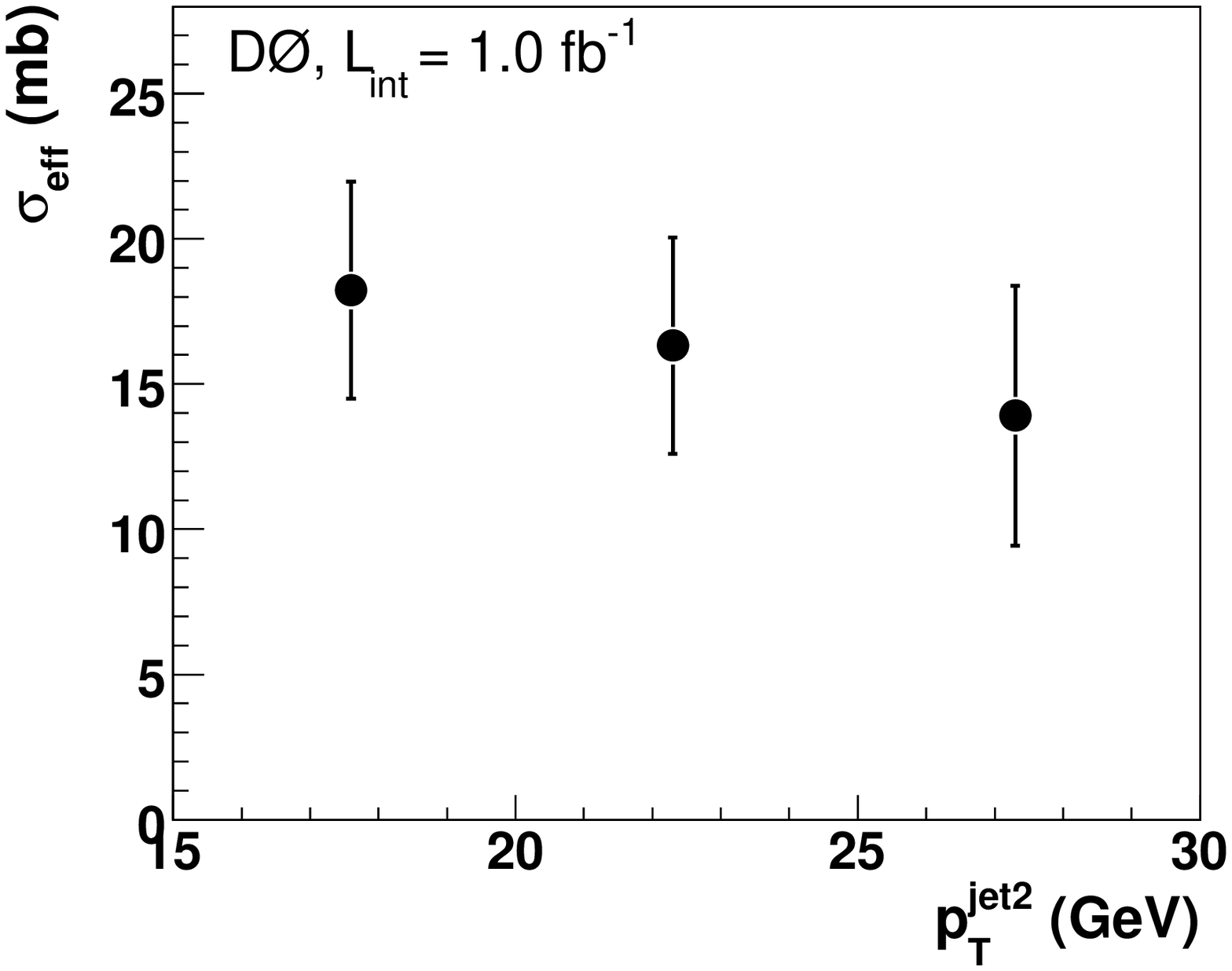}
\vspace*{-5mm}
\caption{The measured effective cross section vs 2nd jet $p_T$.} 
\label{fig:seff_d0}
%
\vspace*{7mm}
\hspace*{0mm}  \includegraphics[scale=0.3]{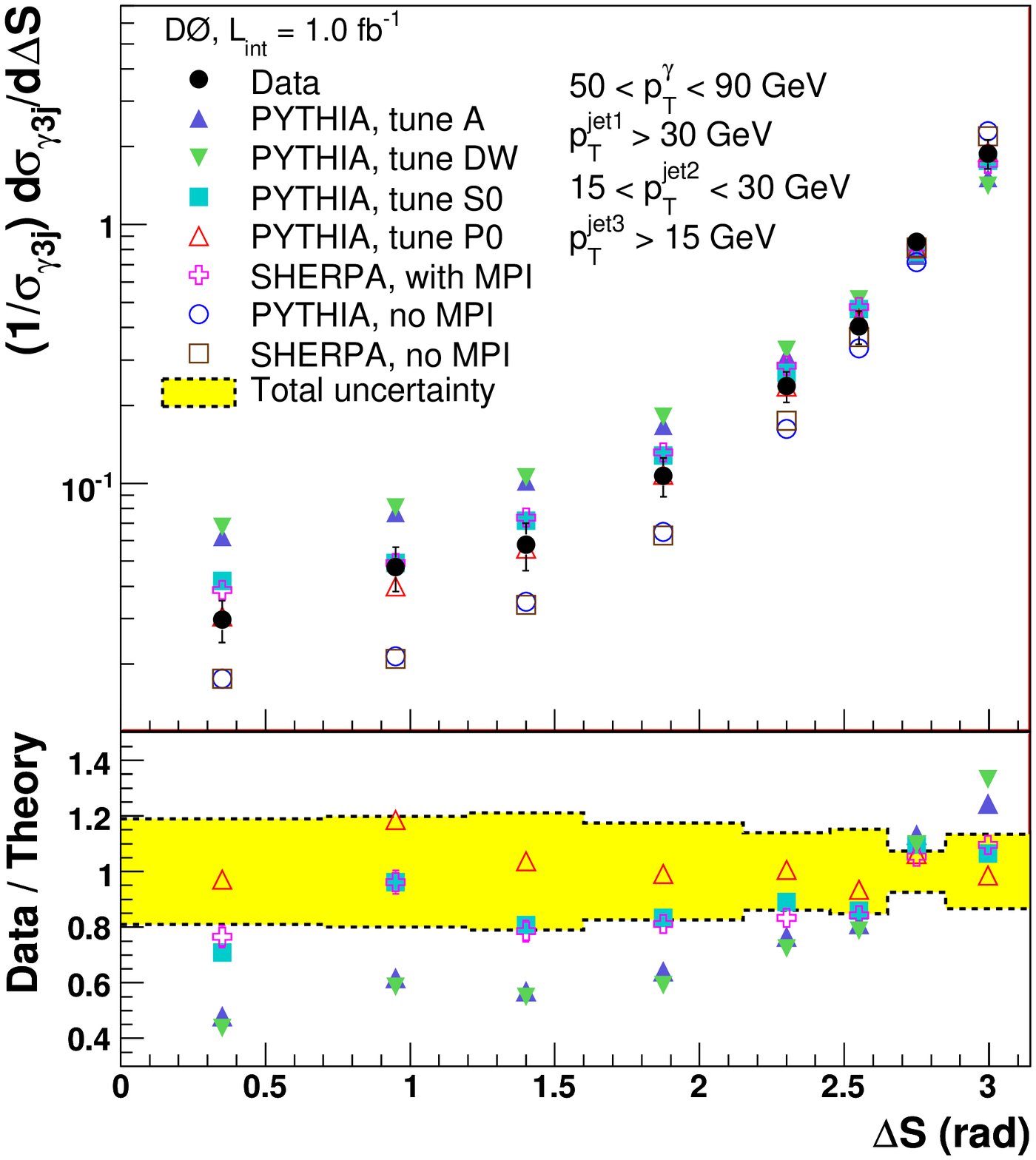}
\hspace*{5mm}  \includegraphics[scale=0.3]{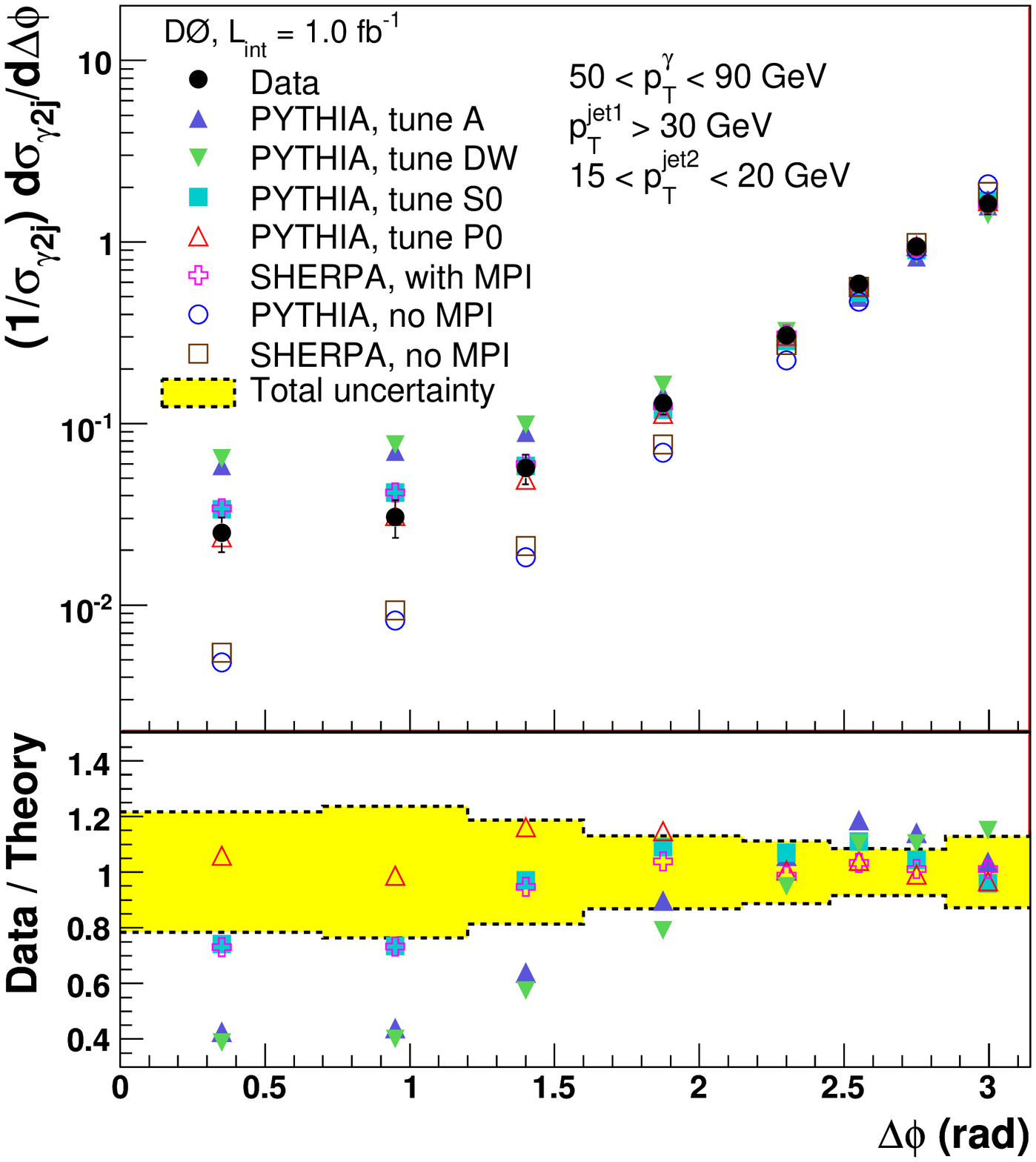}
\caption{Left: Normalized differential cross section in the $\gamma + 3$-jet + X
events, $(1/\sigma_{\gamma 3j})\sigma_{\gamma 3j} /d\Delta S$, in data compared to MC models
and the ratio of data over theory, only for models including
MPI, in the range $15 < p_T^{jet2} < 30$ GeV. 
Right: Normalized differential cross section in $\gamma + 2$-jet + X
events, $(1/\sigma_{\gamma 2j})\sigma_{\gamma 2j} /d\Delta\phi$, in data compared to MC models
and the ratio of data over theory, only for models including
MPI, in the range $15 < p_T^{jet2} < 20$ GeV.} 
\label{fig:ds_d0}
\end{figure}

D0 has studied events with double parton (DP) scattering in $\gamma+3$ jet events \cite{mpi1_d0},
in which two pairs of partons undergo two hard interactions in a single $p\bar{p}$ collision.
The DP events can be a background to many rare processes 
but they also provide insight into the spatial distribution of partons in the colliding hadrons. 
D0 measured the so-called effective cross section, that characterizes rates
of the DP events, $\sigma^{\gamma j, jj}_{\rm DP} = \sigma^{\gamma j}\sigma^{jj} /\sigma_{\rm eff}$.
The measurement was done in the three bins of the 2nd (ordered in $p_T$) jet $p_T$.
The results are shown in Fig.~\ref{fig:seff_d0}. Using these three (almost uncorrelated0 points,
the obtained average effective cross section
is $\sigma_{\rm eff}=16.4\pm 0.3({\rm stat})\pm2.3({\rm syst})$.
It is 
 in agreement with the previous CDF result \cite{mpi_cdf_run1}.
To tune MPI models, D0 also measured cross sections for the azimuthal angle defined between 
the $p_T$ vectors of the $\gamma+$jet and dijet systems in the three $p_T$ bins of the 2nd jet $p_T$ \cite{mpi2_d0}.
Comparison of data with a few MPI and two ``no MPI'' models are shown in Fig.~\ref{fig:ds_d0}.
One can see that data clearly contain DP events and favor more Perugia MPI tunes. \\[-5mm]

\section{Summary}
The Tevatron experiments provide precision QCD measurements of many fundamental observables.
In most cases, the results are mutually consistent and/or complementary to each other.
Jet measurements show good agreement with pQCD, sensitivity to PDF sets,
the strongest constraint on high-$x$ gluon PDF, provide detailed studies 
of different jet algorithms, are used to extract $\alpha_s$, study jet substructure, and
provide limits on many new phenomena models. The $W/Z+$jets results provide  
 extensive tests of pQCD and tune existing MC models.
The photon results test fixed order NLO pQCD predictions accounting for resummation and 
fragmentation effects and show that the theory should be better understood.
Measurements of underlying/MPI events impose strong constraints and improve 
phenomenological MPI models at low and high $p_T$ regimes. 

\begin{footnotesize}


\end{footnotesize}


\end{document}